\documentclass[prb,twocolumn,showpacs,amstext,amsxtra,amssymb,latexsym,bbm,amsmath]{revtex4}
\usepackage{graphicx,epsf}
\usepackage{bm,bbm}

\newcommand{\be}{\begin{equation}}
\newcommand{\ee}{\end{equation}}

\newcommand{\vv}{{\boldsymbol v}}

\newcommand{\p}{{\boldsymbol p}}

\newcommand{\A}{{\boldsymbol A}}
\newcommand{\HH}{{h}}
\newcommand{\ep}{\epsilon}

\def\wideparen#1{\overset{\;\rotatebox{90}{)}}{#1}}

\begin{document}

  \title{Semiclassical quantization of skipping orbits }
  \author{G. Montambaux}
\affiliation{Laboratoire de Physique des Solides, CNRS UMR 8502,
Universit\'e Paris-Sud, 91405- Orsay, France}
\date{\today}

\begin{abstract}
We propose a simple description of the spectrum of edge states in the quantum Hall regime, in terms of semiclassical quantization of skipping orbits along hard wall  boundaries, ${\cal A}=2 \pi (n+\gamma) \ell_B^2$, where ${\cal A}$ is the area enclosed between a skipping orbit and the wall and $\ell_B$ is the magnetic length. Remarkably, this description provides an excellent quantitative agreement with the exact spectrum.
We discuss the value of $\gamma$ when the skipping orbits touch one or two edges, and its variation when the orbits graze the edges and the semiclassical quantization has to be corrected by diffraction effects. The value of $\gamma$ evolves continuously from $1/2$ to $3/4$. We calculate the energy dependence of the drift velocity along the different Landau levels. We compare the structure of the semiclassical cyclotron orbits, their position with respect to the edge, to the wave function of the corresponding eigenstates.
\end{abstract}

  \maketitle

\bigskip

\section{Introduction}

 \label{intro}

The edge states play a crucial role for understanding the integer and
fractional quantum Hall effects. Their description has been
introduced in the seminal paper by  Halperin. \cite{Halperin82,MacDonald84}  This  picture has
then been elaborated by
 Buttiker \cite{Buttiker88} and a review can be found in ref.
[\onlinecite{review}]. It appears often convenient to picture qualitatively these edge states in terms of skipping cyclotron orbits. But the link between the full quantum mechanical treatment of the states and this qualitative picture is missing (see however ref. [\onlinecite{Beenakker89}]). Here we propose an extensive development of this picture and show how the semiclassical quantization of these orbits leads to a qualitative and even quantitative description of the edge states energy
levels.

  We consider a free electron (mass $m$, charge $-e$) moving in a
ribbon infinite  along the $y$ direction and bounded along the $x$
direction.   A magnetic field $B$ is applied along $z$. As in ref. [\onlinecite{Halperin82}], we consider the situation where the confining potential consists in an abrupt potential well of infinite height. This is known not to be the correct situation in the two-dimensional electron gas of $GaAs$ heterostructures, where the confining potential is rather smooth at the scale of the magnetic length $\ell_B$. However, we believe that the case of the abrupt potential is interesting in itself and may be relevant to other related situations exhibiting edge states. For example, graphene ribbons have sharp boundaries which must be modeled with sharp potentials.\cite{BreyFertig,Delplace} We comment the case of smooth boundaries at the end of the paper. Far from the
edges, the energy levels are given by $E_n=(n+1/2) \hbar \omega_c$
where $\omega_c=e B/m$ is the cyclotron frequency. This Landau
quantization can be obtained quite easily from the Bohr-Sommerfeld
quantization rule that we recall below. The goal of this work is to
describe the semiclassical motion of the electron near one edge,
described here by a "hard wall", that is an infinite potential well. We consider the
vicinity of the edge located at $x=0$, assuming first that the
second edge is far away ($x \rightarrow -\infty$).

The problem is solved semiclassically by quantization of the action.
In the appropriate gauge, the well-known Landau problem is related
to the problem of a one-dimensional oscillator. In the presence of
a sharp edge, the problem to be
solved is  the one of an harmonic oscillator in the presence of an infinite
potential well. This problem is solved by quantization of the
semiclassical action $S(E)=2 \pi (n+ \gamma) \hbar$, where $\gamma$
is related to a so-called Maslov index.\cite{Maslov} For the free oscillator,
$\gamma=1/2$ corresponds to the sum of two contributions $ \gamma_i=1/4$ of
the two turning points. In the presence of the potential well, $\gamma$
evolves from $1/2$ to $3/4$ when the guiding center $x_c$ of the
cyclotron orbit (the center of the harmonic oscillator) approaches
the wall. A form of the continuous variation $\gamma_n(x_c)$ for a
given $n$ has been recently obtained.\cite{Avishai08}

In this paper, we give a very simple description of the edge states spectrum in
terms of quantization of skipping orbits. This image, currently used
in the literature or in pedagogical presentations, has curiously
never been described in details (see however ref. [\onlinecite{Beenakker89}]). Yet, it leads to
a number of results which to our knowledge have never been
discussed. In the next section, we recall the mapping, in the Landau
gauge, to a one-dimensional problem of a harmonic oscillator and we
calculate the action $S(E)$ of this oscillator. In section \ref{sect.3}, we
give a complete   picture of the evolution of the energy levels in
terms of the quantization of the area of skipping orbits. The well-known quantization of closed orbits can be extended to the case of skipping orbits. Then, their area depends on the distance $x_c$ to the wall and  must be quantized as
\be {\cal A}(R,x_c)= 2 \pi (n + \gamma) \ell_B^2  \label{QA}  \ee
where $\ell_B$ is the magnetic length, $R$ is the cyclotron radius and $x_c$ is the position of the guiding center with respect to the wall. This well-known quantization rule for closed orbits appears to be also valid for open but periodic skipping orbits. Using the same method, we calculate in section \ref{sect.4} the full spectrum in the case a ribbon, when the magnetic length is of the order of the width of the ribbon so that the two edges have to be considered.
Then we conclude in section \ref{sect.5}, with a comparison with the case of a smooth potential.

\section{Mapping to a one-dimensional oscillator}
\label{sect.2}

\begin{figure}[!h]
\centerline{ \epsfxsize 8cm \epsffile{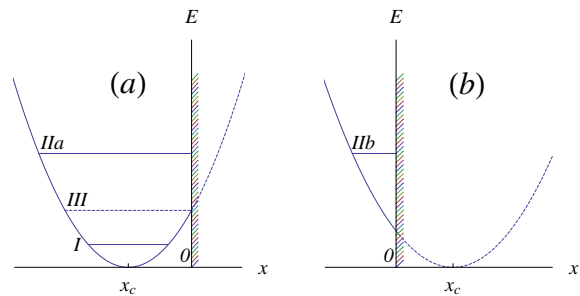} } \caption{\it
$1D$ harmonic oscillator centered on $x_c$, with an infinite
potential well in $x=0$. a) The guiding center is inside the sample. b) The guiding center is outside the sample. Three distinct regions have to be considered : I : the energy levels are not affected by the edge and $\gamma=1/2$, II: the trajectories hit the edge and $\gamma=3/4$. In between, the phase factor $\gamma$ evolves between $1/2$ and $3/4$.  It is $2/3$ when the  cyclotron orbit just grazes the wall (III). }   \label{regimes}
\end{figure}

 The problem to be solved is described by the Hamiltonian
 \be {\cal H}= {p_x^2 \over 2 m}+ {p_y^2 \over 2 m} + V(x) \ee
 where the potential $V(x)$ describes the edge of the sample along the $y$ direction. We choose
 $V(x)=0$ when $x <0$ and $V(x)= \infty$ when $x >0$. Using the Landau gauge $\A=(0, Bx,0)$, the
  corresponding eigenvalue problem   reads ($-\infty < x \le 0$)~:
\be [- {\hbar^2 \over 2 m}{d^{2} \over dx^{2}}+{1 \over 2} m
\omega_c^2 (x_{c}-x)^{2} ] \psi(x)=E \psi(x),     \label{SE}
 \ee
with the constraint  $\psi(0)=0$. The center $x_c$ of the
oscillator,
 is related to the $k_y$ component of the wave vector which is a
good quantum number: $x_c=k_y \ell_B^2$  where $\ell_B=\sqrt{\hbar /
e B}$ is the magnetic length. The action $S$ along a closed trajectory    is given by $ S=   \oint
p_x dx= 2 \int
p_x dx$ where, far from the edge (region I in Fig. \ref{regimes})
\cite{factor2}
\begin{eqnarray} S&&=2 \int_{x_c-R}^{x_c+R} \sqrt{2 m E - m^2 \omega_c^2
(x-x_c)^2} dx \nonumber \\
&&=
 {2 \pi E \over \omega_c} = {\pi } m \omega_c R^2. \label{actionI}
 \end{eqnarray}
 $R(E)=\sqrt{2 E / m \omega_c^2}$  is the cyclotron radius of the classical
trajectory and $x_c \pm R$ are  the positions of the
turning points. The semiclassical quantization of the action
 \be  S(E)=2 \pi (n + \gamma) \hbar \label{qrule1} \ee
leads to  the energy quantization,
 \be E=(n+\gamma)\hbar \omega_c \label{Egamma} \ee
where the value $\gamma=1/2$ is not given by the semiclassical quantization rule and  results from the matching  of the wave
function at the two turning points.

 \bigskip

When the cyclotron orbit approaches the edge, that is when the
center $x_c$ of the cyclotron orbit becomes larger than $-R$ (energy
regions
  IIa and IIb   in Fig. \ref{regimes}), the turning points are located at
  $x_1=x_c-R$ and $x_2=0$. The
 action now explicitly  depends on $x_c$ and it is given by \cite{factor2}
\begin{eqnarray}
  S(E,x_c) &=& 2 \int_{x_c-R}^{0} \sqrt{2 m E - m^2 \omega_c^2 (x-x_c)^2} dx
  \nonumber \\
  &=& {2 E \over  \omega_c} [ {\pi \over 2} -
 \arcsin {x_c \over R} - {x_c \over R  } \sqrt{1 - x_c^2/R^2} ] .               \label{actionII}
 \end{eqnarray}
 Introducing the angle $\theta$ such that $\cos \theta=x_c/R$, the
 action can be rewritten as
\be S(E,x_c)= {E \over  \omega_c}[2 \theta - \sin 2 \theta] = {1
 \over 2} m \omega_c R^2 [2 \theta - \sin 2 \theta] \ .  \label{qrule2} \ee
  We give in the next section a very
simple interpretation of this angle $\theta$. The energy levels  $E_n(x_c)$ are still given by quantization of the action (\ref{qrule1}) which now depends on the position with respect to the wall. When $x_c > -R$, the factor $\gamma$ is equal to $3/4$, because it results from different matching conditions at the two turning points. At the left turning point $\gamma_l=1/4$, while at the right turning point, the vanishing of the wavefuntion implies $\gamma_r=1/2$, so that $\gamma=\gamma_l+\gamma_r=3/4$. $\gamma$ evolves between $1/2$ and $3/4$ when $x_c \simeq - R$.

\begin{figure}[!h]
\centerline{ \epsfxsize 7cm \epsffile{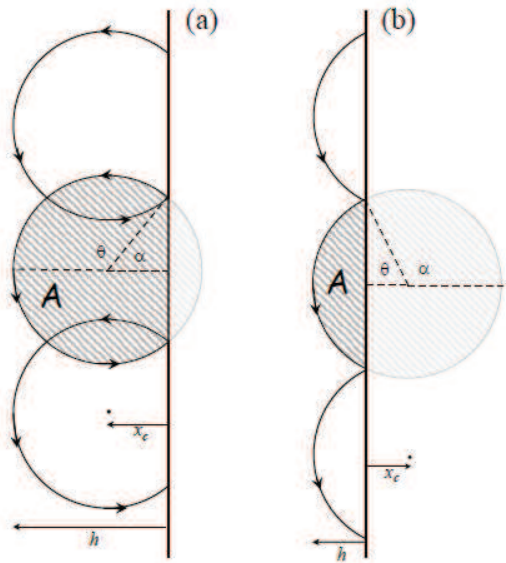} } \caption{\it
Semiclassical skipping orbits a) when the guiding center   is
inside the sample (region IIa in Fig. \ref{regimes}); b) when it is outside (region IIb in Fig. \ref{regimes}). The dashed area ${\cal A}$ is quantized as ${\cal A}= 2 \pi (n+\gamma) \ell_B^2$.} \label{fig.skipping}
\end{figure}

\section{Quantization of skipping orbits}
\label{sect.3}
\subsection{Quantization of the area}

\begin{figure}[!h]
\centerline{ \epsfxsize 7cm \epsffile{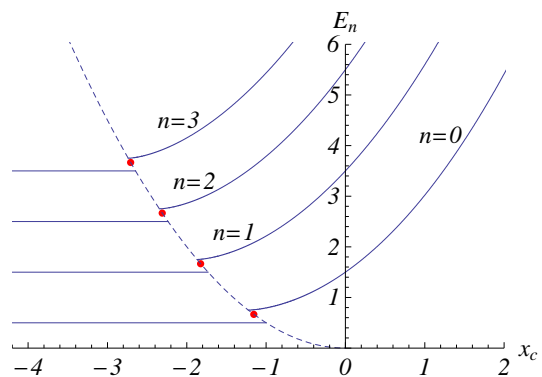} }
\caption{\it Energy levels obtained from semiclassical quantization
of the area ${\cal A}(E,x_c)= 2 \pi (n+ \gamma) \ell_B^2$. Inside
the sample ($x_c <<  -R$), $\gamma=1/2$. For skipping orbits ($x_c
>> -R$), $\gamma=3/4$. The dashed line corresponds to $x_c=-R$, the case where the cyclotron orbit just touches the
edge. In this case, we have found that $\gamma=2/3$ and the position
of the energy levels is marked with small dots along the dashed
line. The energy levels are plotted in units of $\hbar \omega_c$ and the distance $x_c$ is plotted in units of $\ell_B$.} \label{spectresimple}
\end{figure}

The quasiclassical Bohr-Sommerfeld quantization rule for skipping orbits has been discussed by Beenakker and Van Houten \cite{Beenakker89}. Since the motion along the $x$ axis is periodic, this quantization rule can be written as

 \be  S= \oint p_x \cdot d x = 2 \pi (n+\gamma) \hbar  \ . \label{qrule3} \ee
where $\p= m \vv - e \A$. The trajectories are now open and the notation $\oint$  means that the integral is taken along one period of the motion.  The gauge for the vector potential $\A$  must be chosen such that $A_x$ is periodic. The simplest choice is $A_x=0$,
so that $p_x=m v_x$.  The classical equation of motion for the $x$ component of the velocity  is $v_x=-\omega_c (y - y_0)$ where $y_0$ is an arbitrary position.
Therefore the
 quantization condition (\ref{qrule3}) becomes

\be m \oint v_x d x= - e B  \oint (y - y_0) d x= 2 \pi (n+\gamma) \hbar  \ .  \label{qrule4} \ee
The integral is nothing but the  area ${\cal A}$ enclosed between one arc of the periodic orbit and the wall (see Fig. \ref{fig.skipping}). Therefore we can generalize the familiar quantization rule (\ref{QA}) of the area ${\cal A}$ to the case of skipping orbits. This area can be parametrized by the angle $\theta$
shown in Fig. \ref{fig.skipping} and defined by $x_c=R \cos \theta$. We have
\be {\cal A}(E,x_c)= {R^2 \over 2 }[2 \theta - \sin 2 \theta]
\label{Areatheta} \ee
 which is precisely the same equation (\ref{qrule2}) as obtained in the 1D picture. Then the  quantization of the area
  ${\cal A}$ reads
\be {\cal A}(R,x_c)= 2 \pi (n + \gamma) \ell_B^2 \label{Arean} \ee
so that the angle $\theta$ can be used to parametrize the solutions
($\theta=\pi$: the orbit just grazes the edge, $x_c=-R$. $\theta=\pi/2$: the
guiding center of the orbit in precisely on the edge. $\theta <  \pi/2$: the center
of the orbit  stands outside the sample). From equations
(\ref{Areatheta},\ref{Arean}), we obtain

\be R^2 = {4 \pi (n+\gamma) \ell_B^2 \over 2 \theta - \sin 2 \theta
} \label{R2} \ee
and the energy levels are given by
 \be E=\hbar \omega_c {R^2 \over 2
\ell_B^2}= \hbar \omega_c (n+\gamma) {2 \pi \over 2 \theta - \sin 2
\theta } \label{E1} \ee
The cyclotron radius $R$ is related to the position $x_c$ of the
guiding center:
\be x_c= R \cos \theta =  \ell_B \sqrt{4 \pi (n+\gamma)  \over 2
\theta - \sin 2 \theta }  \cos \theta  \label{RR1}\ee
so that the dependence $E_n(x_c)$  is simply parametrized by the
angle $\theta$. However, the main complexity of the problem comes
from the fact that  $\gamma$ is not a constant. It is
fixed to the value $1/2$ far from the edge, but on the other hand,
for skipping orbits, it reaches the value $3/4$. Therefore, from the
quantization condition (\ref{Arean}), we obtain two branches (Fig. \ref{spectresimple}).

\subsection{Spectrum}

In the intermediate region, when the cyclotron orbit is very close
to the wall, that is when $x_c \simeq -R$,   $\gamma$
varies continuously between $1/2$ and $3/4$. This regime has been
studied recently within a WKB approach.\cite{Avishai08} In particular, when the cyclotron
orbit strictly touches the wall $x_c= -R$, it has been found that the
parameter $\gamma=2/3$. This factor comes from a contribution $1/4$
on the left side and a very peculiar and new contribution $5/12$
from the right side, which, to our knowledge has never been studied,
at least in this context. Moreover, in ref. [\onlinecite{Avishai08}], we have found an
interpolation formula for $\gamma_n(x_c)$ for a given value of $n$.
It is given by
\be \gamma_n(x_c)= {1 \over 2}\ {1 + 3 e^{A X} \over 1 + 2 e^{A X}}
\label{gammanxc} \ee
where $X= (2n + 4/3)^{1/6} (x_c/\ell_B + \sqrt{2 n + 4/3})$ and $A
\simeq 3.5$. This expression can be extended  by transforming it into a function of energy and $x_c$ : $\gamma(E,x_c)$ is still given by Eq. (\ref{gammanxc}), with
$X= (2 E/\hbar \omega_c)^{1/6} (x_c/\ell_B + \sqrt{2 E/\hbar \omega_c})$. It can be actually decomposed in the form

\be \gamma(E,x_c)= {1 \over 2}+ \gamma_r(E,x_c)
\label{gammadec} \ee
since it is known to be the contribution of two terms corresponding respectively to the left and to the right turning points. We have

\be \gamma_r(E,x_c)=  {1 \over 4}\ {1 + 4 e^{A X} \over 1 + 2 e^{ A X}}
\label{gammar} \ee

\begin{figure}[!h]
\centerline{ \epsfxsize 7cm \epsffile{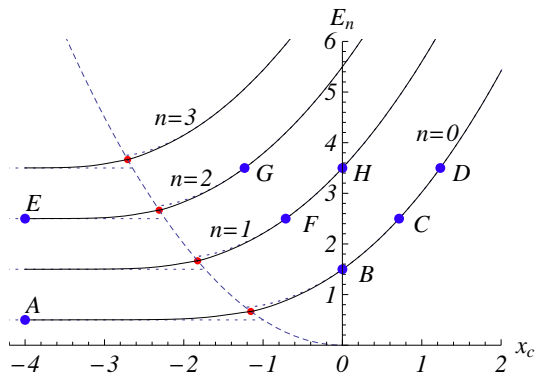} }
\caption{\it Black curves: Energy levels obtained from semiclassical quantization
of the area ${\cal A}(E,x_c)= 2 \pi (n+ \gamma_n(x_c)) \ell_B^2$,
where we have used the expression (\ref{gammanxc}) for
$\gamma_n(x_c)$. The dashed line corresponds to $x_c=-R$, the case
where the cyclotron orbit just touches the edge. On this line,
$\gamma=2/3$. The dotted lines are the approximation of a constant $\gamma=1/2$ or $3/4$, see Fig. \ref{spectresimple}. The large dots indicate special points where the wave
function can be easily obtained from the solution of the harmonic
oscillator in free space (see Figs. \ref{fondamental},\ref{n2}).
When $x_c \simeq -R$, $\gamma$ varies continuously between $1/2$ and
$3/4$ so that the spectrum is continuous. The energy levels are plotted in units of $\hbar \omega_c$ and the distance $x_c$ is plotted in units of $\ell_B$.} \label{spectrecomplet}
\end{figure}

Given these expressions and the implicit equation
(\ref{Arean}), the full spectrum is obtained in Fig. \ref{spectrecomplet}.

The scenario when the cyclotron orbits approaches the edge is the
following. When $x_c \ll -R$, that is far from the edge, the
cyclotron radius is $R=\ell_B \sqrt{2 n +1}$. When the distance
between the orbit and the edge becomes of order of the magnetic
length $\ell_B$, the energy and the cyclotron radius start to
increase to reach the values $E_n=(n+2/3) \hbar \omega_c$ and
$R=\ell_B \sqrt{2 n +4/3}$ when the cyclotron orbit just touches the
edge. Then the energy and the cyclotron radius continue to increase
as shown in Fig. \ref{spectrecomplet}.

\begin{figure}[!h]
\centerline{ \epsfxsize 7cm \epsffile{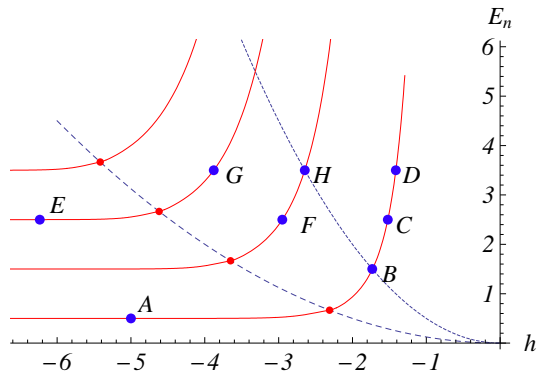} }
\caption{\it  Energy levels $E_n(\HH)$ obtained from semiclassical
quantization. $\HH$ is the position of the extremum of the cyclotron
orbit. The dashed line corresponds to $\HH=- 2 R$, the case where the
cyclotron orbit just touches the edge. On this line, $\gamma=2/3$.
The dotted line corresponds to $\HH=-R$, that is $x_c=0$. The large
dots indicate special points where the wave function can be easily
obtained from the solution of the harmonic oscillator in free space
(see figures \ref{fondamental},\ref{n2}).   The energy levels are plotted in units of $\hbar \omega_c$ and the distance $\HH$ is plotted in units of $\ell_B$}
 \label{spectreh}
\end{figure}

It is also interesting to introduce the position $\HH$ of the extremum
of the cyclotron orbit (Fig. \ref{fig.skipping}), that is the position
of the left turning point in the $1D$ picture. It is $\HH=-2 R$ when
the cyclotron orbit just touches the edge and it varies to $\HH
\rightarrow 0$ when $x_c \rightarrow \infty$. A simple geometric
picture shows that $\HH=x_c-R=R (\cos \theta -1) <0$, that is, using
\ref{R2}:

 \be \HH=R(
\cos \theta-1) = \ell_B \sqrt{4 \pi (n+\gamma) \over 2 \theta - \sin
2 \theta }  (  \cos \theta-1) \ee
In Fig. \ref{spectreh}, we plot the energy as a function of the
position $\HH$. Of course, $x_c$ can increase to infinity and $\HH$ stays
confined to the inside of the sample ($\HH <0$).  (cf. inflexion
point).

\begin{figure}[!h]
\centerline{ \epsfxsize 7cm \epsffile{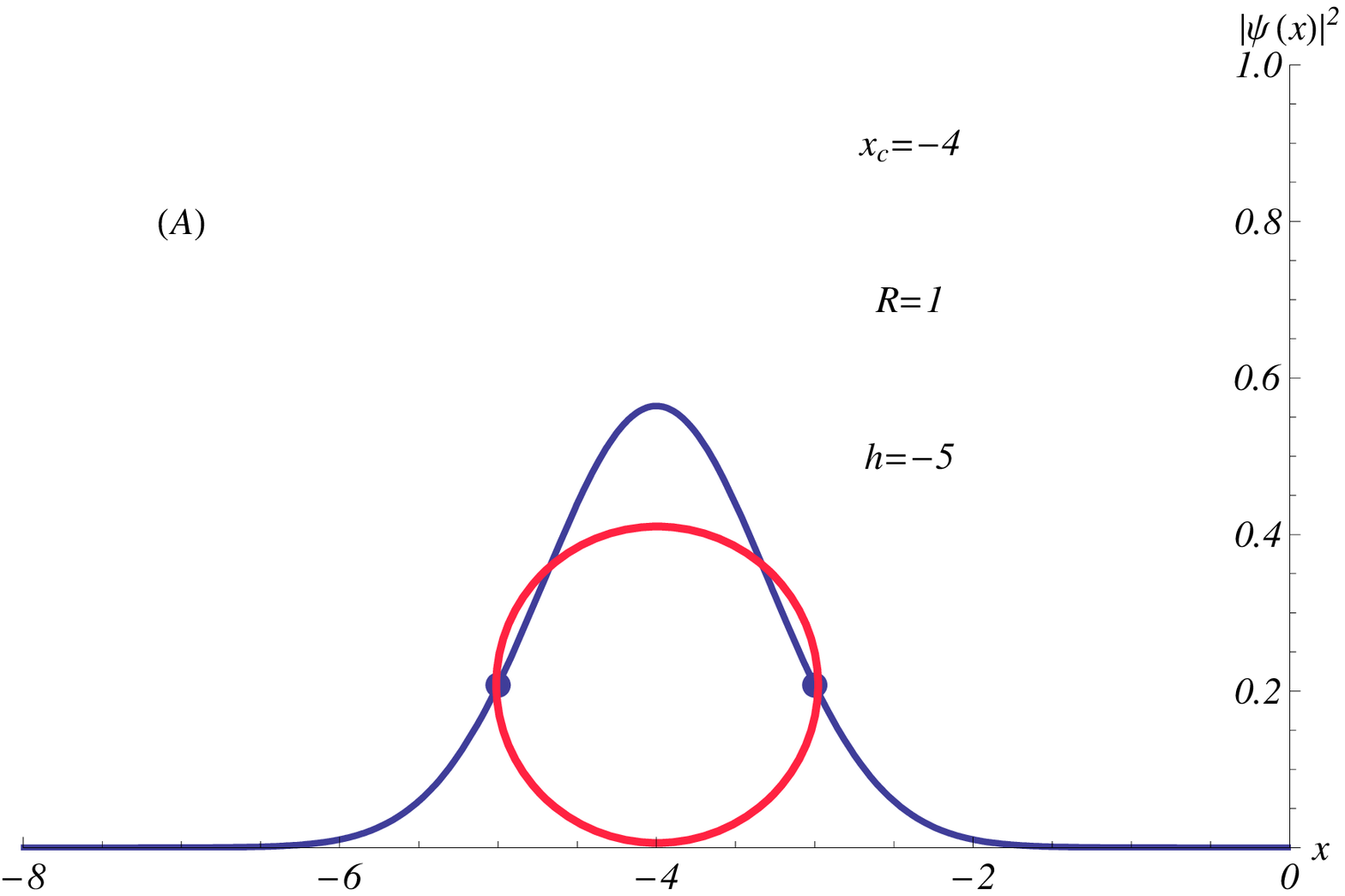} }
\centerline{ \epsfxsize 7cm \epsffile{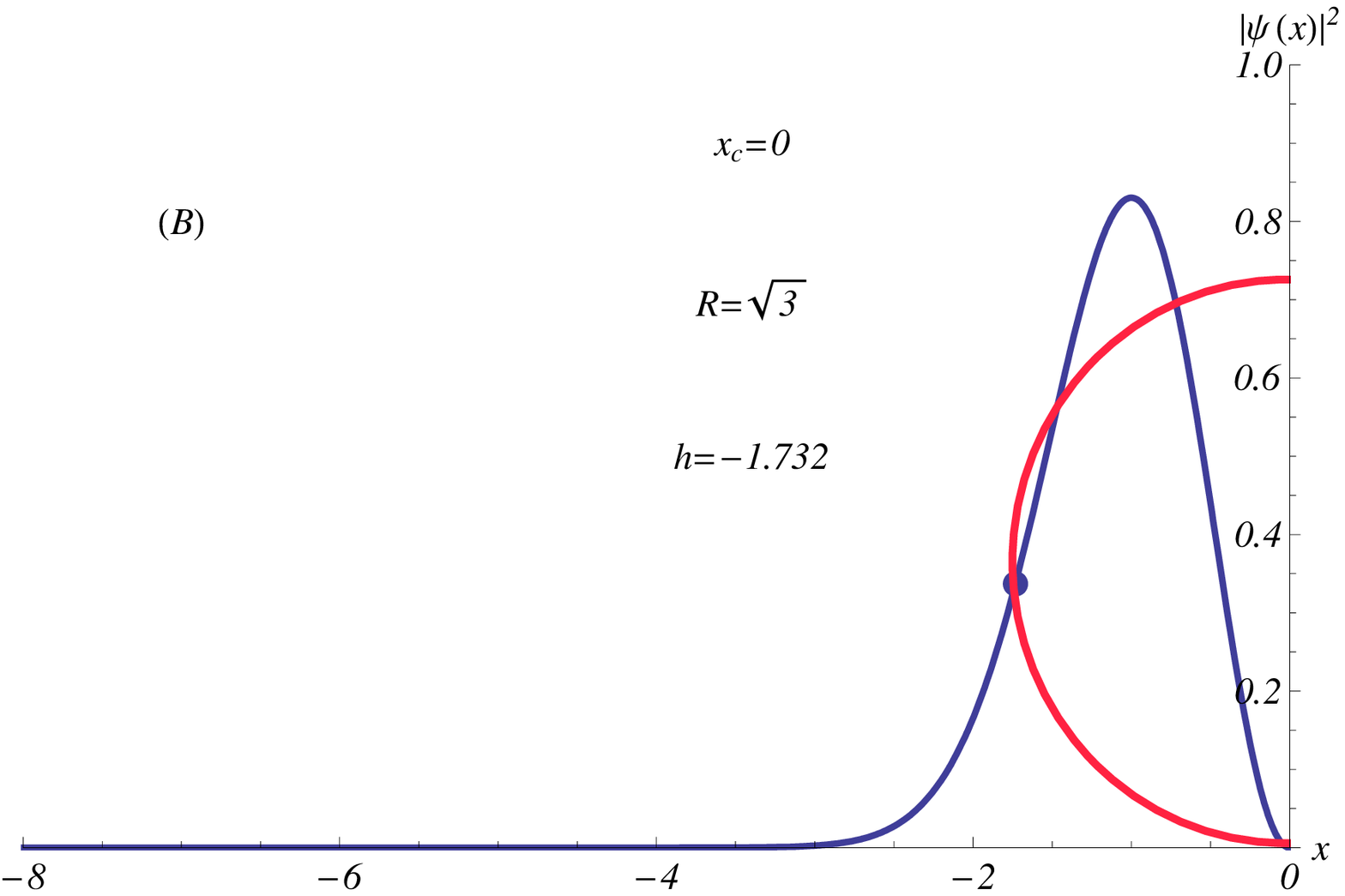} }
\centerline{ \epsfxsize 7cm \epsffile{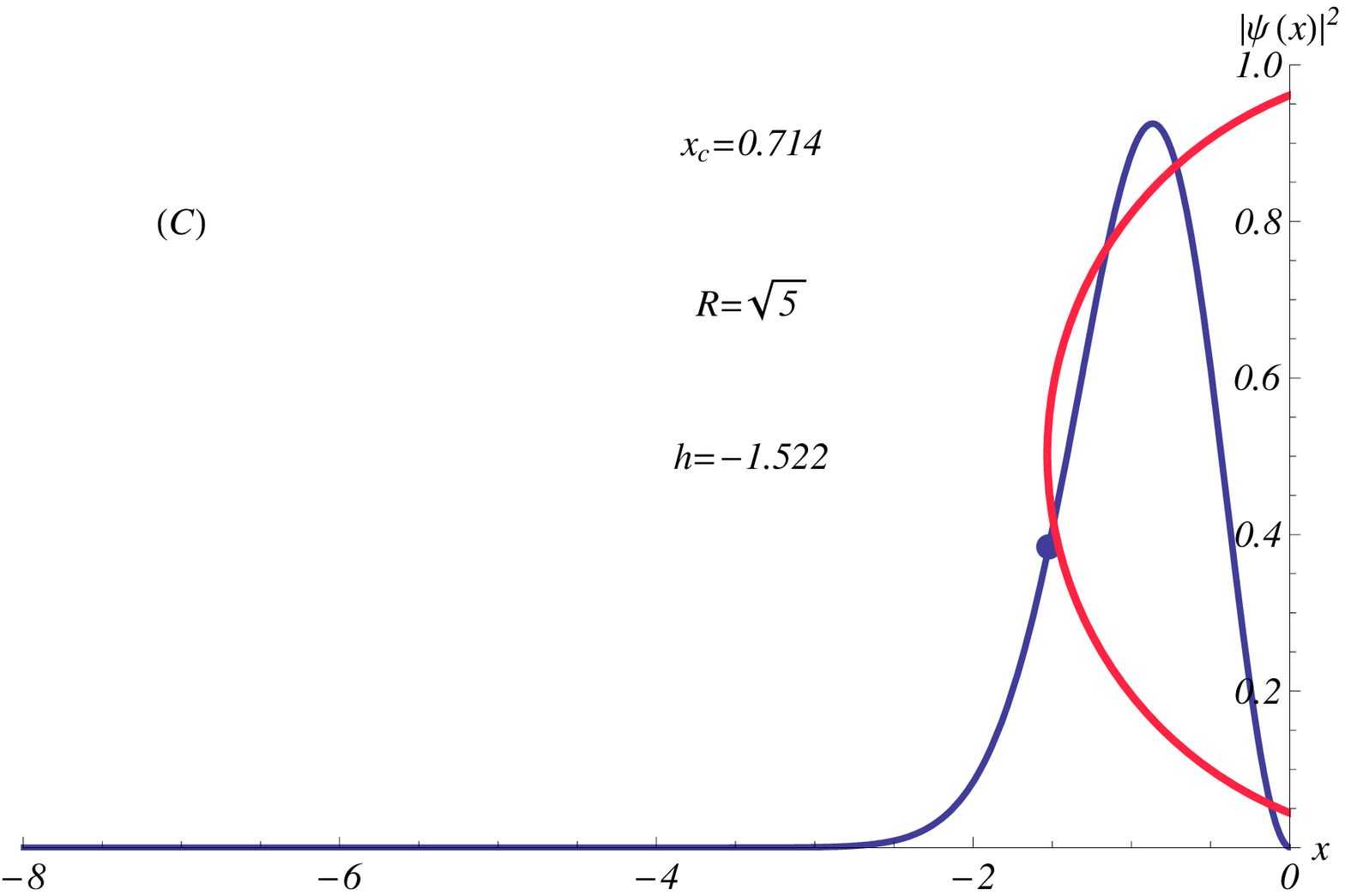} }
\centerline{ \epsfxsize 7cm \epsffile{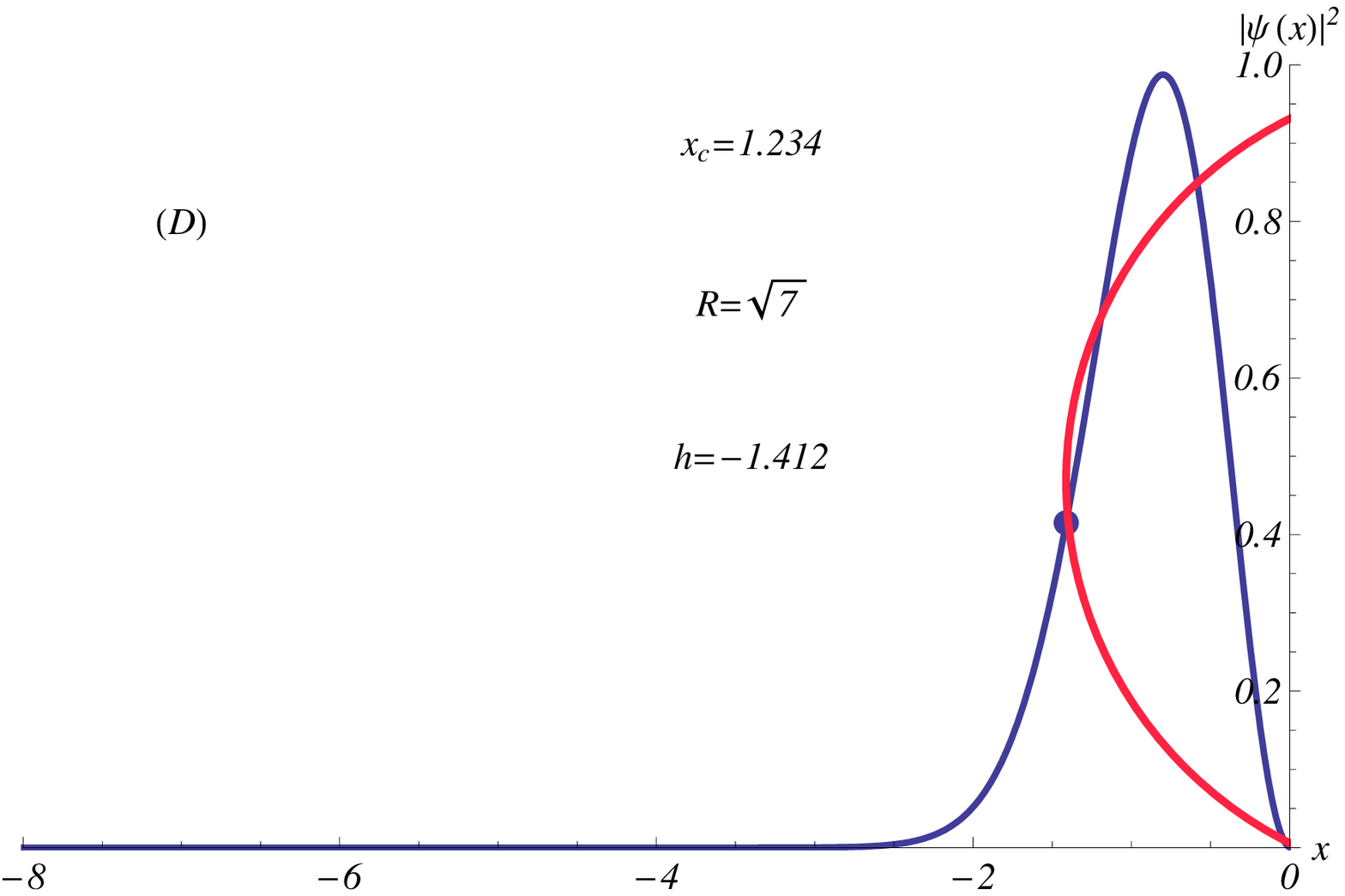} }
\caption{\it Evolution of the ground state wave function
($n=0$) when the position $x_c$ of the guiding center approaches and
crosses the edge. The four cases correspond  to the points (A), (B), (C)  and (D) shown in Figs. \ref{spectrecomplet},\ref{spectreh}. The position of the turning point (at distance
$|\HH|$ from the edge) is marked with a dot.   Is is seen that the extension of the wave function is given by the extremum of the classical skipping orbit. The distance $x$ is plotted in units of $\ell_B$. } \label{fondamental}
\end{figure}

\begin{figure}[!h]
\centerline{ \epsfxsize 7cm \epsffile{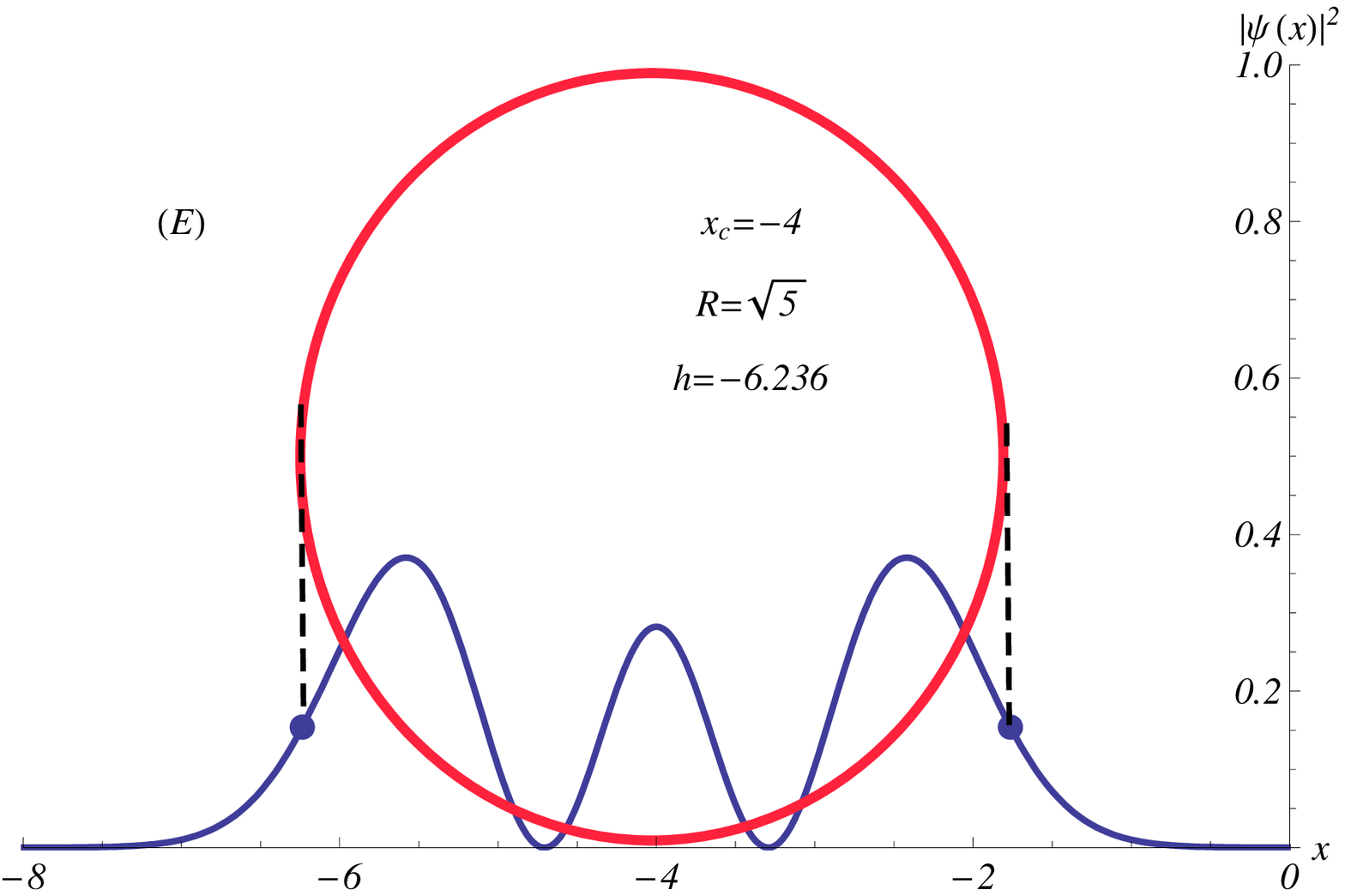} }
\centerline{ \epsfxsize 7cm \epsffile{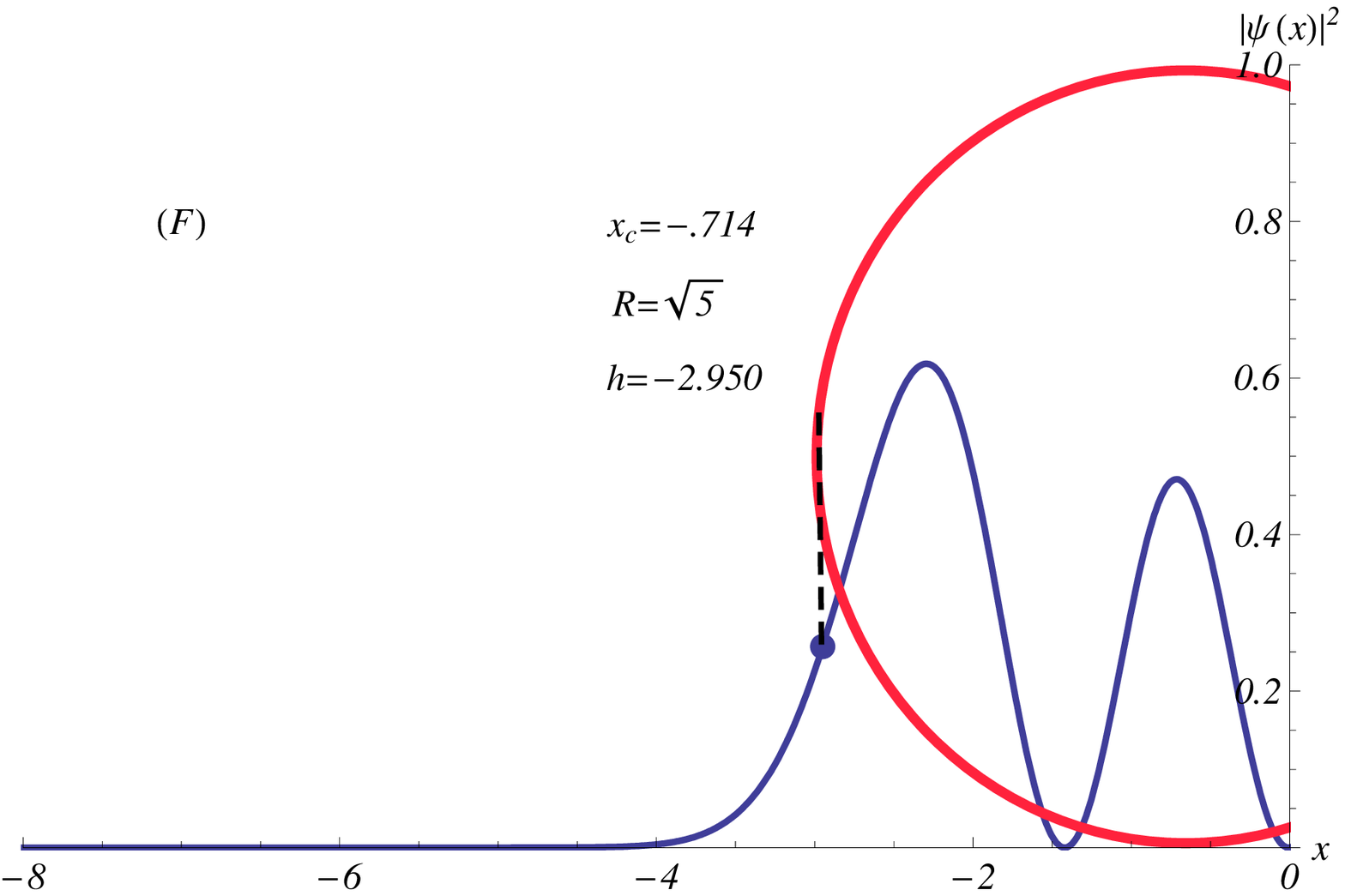} }
\centerline{ \epsfxsize 7cm \epsffile{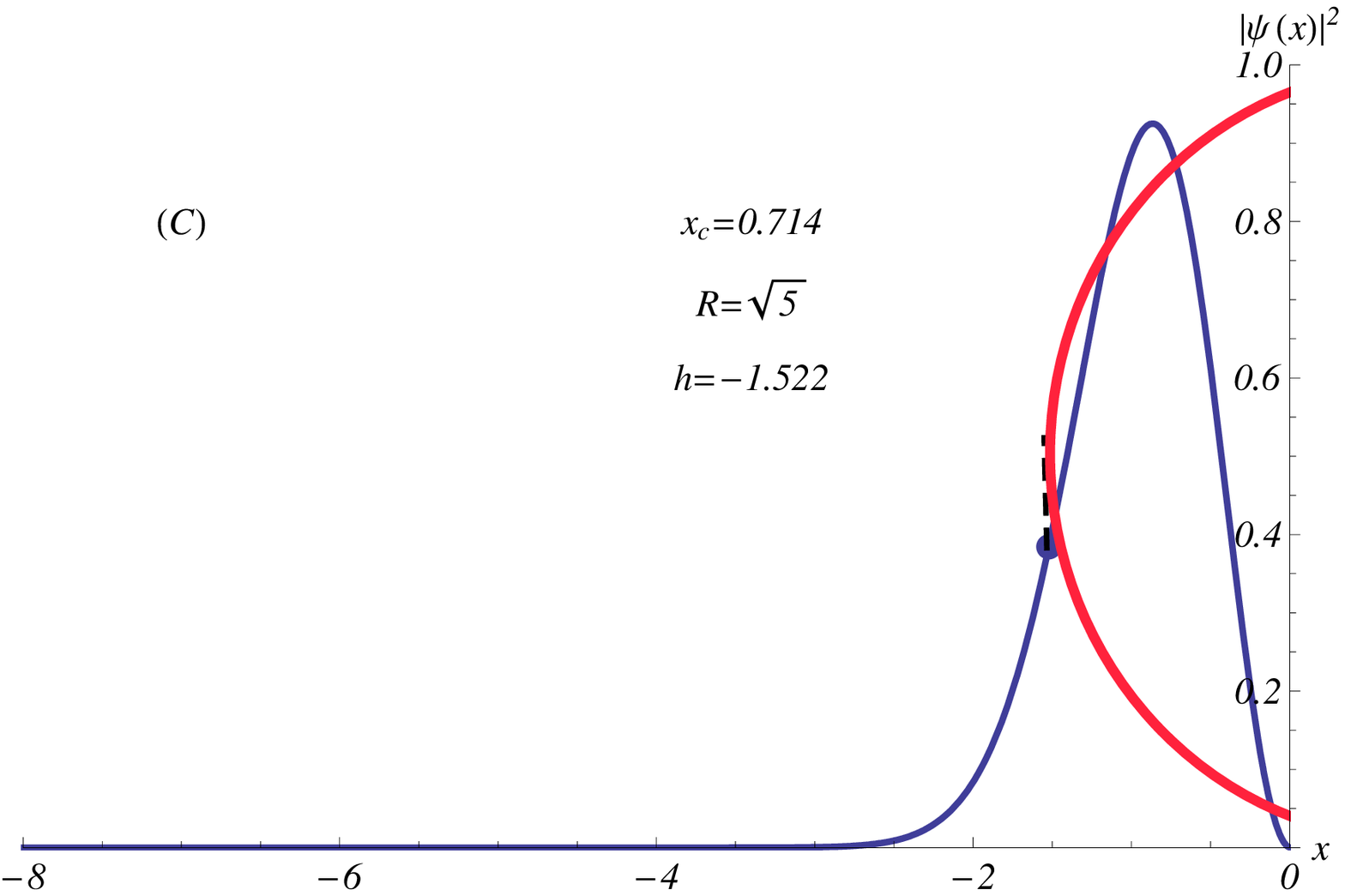} }
\caption{\it
Evolution of the state of energy $5/2 \hbar   \omega_c$  when $x_c$ increases,
corresponding to the points (E), (F) and (C) shown in Figs. \ref{spectrecomplet},\ref{spectreh}. The extension of the wave function is given by the extremum of the semiclassical orbit. Distances are   in units of $\ell_B$.} \label{n2}
\end{figure}

The large dots marked in Figs. \ref{spectrecomplet},\ref{spectreh}
correspond to simple cases where the energy and the wave function
are easily known. For these special points, where the energy is  the
same as in free space ($3/2, 5/2, 7/2 \times \hbar \omega_c$), the wave function is also
the same as in free space but must vanish in $x=0$. The wave
functions in free space are well known to be related to the Hermite
functions. Therefore the edge must coincide with a zero of these
Hermite functions. For example the points $B$ and $H$ correspond to
antisymmetric wave functions, that is to energies $E_n=E_{2 p +1}=
\hbar \omega_c (2 p + 3/2)$.  In Fig. \ref{fondamental}, we have
shown the evolution of the normalized squared wavefunction
$|\psi(x)|^2$ for increasing values of $x_c$. In Fig. \ref{n2},
we show three wave functions with energy $5/2 \hbar \omega_c$, having respectively
$two$, $one$ and no zeroes. In these two figures, one sees that the extension of the wave function is given by the extremum of the classical skipping orbit.

\subsection{Drift velocity}

\begin{figure}[!h]
\centerline{ \epsfxsize 7cm \epsffile{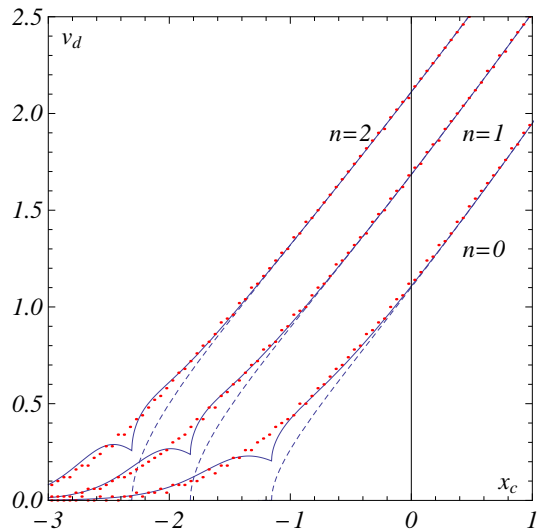} } \caption{\it
Variation of the drift velocity $v_d$ (in units of $\hbar/m\ell_B$) with the position $x_c$ for the three lowest energy levels. Red dots: exact numerical calculation; full line: result (\ref{vdWKB}) of the WKB calculation; dashed line: classical drift velocity (\ref{vdcl}). $x_c$ is in units of $\ell_B$.
 } \label{fig.vdrift1}
\end{figure}

We now calculate semiclassically the drift velocity along an edge state at a given energy $E$. The cyclotron radius is $R= \sqrt{2 E / m \omega_c^2}$ and the velocity along the cyclotron orbit is  given by $v_E=\omega_c
R=\sqrt{2 E/m}$. The length of the skipping orbit being $2 R
\theta$ (Fig. \ref{fig.skipping}), the period $T$ is given by
\be T=2 R \theta /v= {2 \theta \over \omega_c} \ .  \ee
Classically, the drift velocity $v_d^{cl}$ along the $y$ direction can be easily
obtained from simple geometry (Fig. \ref{fig.skipping}), the distance between two successive hits on the
edge being $2 R \sin \theta$:
\be v_d^{cl}={2 R \sin \theta \over T}= \omega_c R {\sin \theta
\over \theta}=v_E {\sin \theta \over \theta} \label{vdcl} \ee
It is interesting to compare this value to the drift velocity
obtained from the energy
\be v_d= {\partial E \over \hbar \partial k_y}={1 \over e B}
{\partial E \over \partial x_c} \label{vdE} \ee
which is the correct result, beyond semiclassical approximation. The derivative can be calculated from Eqs. (\ref{E1}, \ref{RR1}), and one recovers the classical expression (\ref{vdcl}) provided $\gamma$ is a constant. However in the region where the
cyclotron orbit is near the edge $x_c \simeq -R$, $\gamma$ varies
continuously between $1/2$ and $3/4$. A better evaluation of the drift velocity is obtained in the WKB approximation which accounts for the variation of $\gamma$ (Eq. \ref{gammanxc}). The energy levels are given by  the two equations
\be R^2= {4 \pi [n + \gamma_n(x_c)] \over 2 \theta - \sin 2 \theta} \ell_B^2 \qquad, \qquad x_c=R \cos \theta \label{eqnsWKB} \ee
Since $E= \hbar \omega_c R^2/2 \ell_B^2$, the drift velocity
is
\be v_d^{WKB}={1 \over e B } {\partial E \over \partial x_c}= \omega_c R {\partial R \over \partial x_c} \ee
By differentiating Eqs. (\ref{eqnsWKB}), and eliminating $\partial R/\partial \theta$, we obtain
\be v_d^{WKB}= \omega_c R {\sin \theta \over \theta} + 2 \pi \omega_c \ell_B^2 {\sin^2 \theta \over \theta (1-  \cos 2 \theta)} {\partial \gamma_n \over \partial x_c} \ee
The derivative $\partial \gamma_n \over \partial x_c$ is non zero only in the vicinity $x_c \simeq -R$, in which case the angle $\theta$ is very close to $\pi$. Therefore, in a very good approximation, we obtain:

\be v_d^{WKB}= v_d^{cl}+ \omega_c \ell_B^2  {\partial
\gamma_n \over \partial x_c}  \ . \label{vdWKB}  \ee
The dependence $v_d^{WKB}(x_c)$ is plotted in Fig. (\ref{fig.vdrift1}) and is compared to the   numerical fully quantum calculation.
The drift velocity, which
is zero inside the sample, starts to increase when the cyclotron
orbit touches the edge.
The WKB approximation is   excellent   except close to the point where the classical cyclotron orbit just touches the boundary.  When the skipping orbit gets closer and closer to the edge, the energy increases and the classical approximation (\ref{vdcl}) becomes excellent.

\begin{figure}[!h]
\centerline{ \epsfxsize 7cm \epsffile{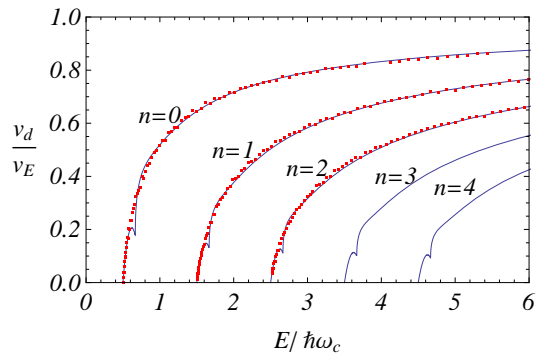} } \caption{\it
Variation of the drift velocity with the energy along different Landau levels. The drift velocity is normalized to the Fermi velocity $v_E$ corresponding to energy $E$. The full lines are the  WKB calculations and the dots are the exact results.} \label{fig.vdrift2}
\end{figure}

Fig. \ref{fig.vdrift2} represents the drift velocity normalized to the Fermi velocity, as a function of the energy along a given edge state. The drift velocity ultimately saturates towards the Fermi velocity at high energy.

.

\section{Two edges}
\label{sect.4}

\begin{figure}[!h]
\centerline{ \epsfxsize 8.5cm \epsffile{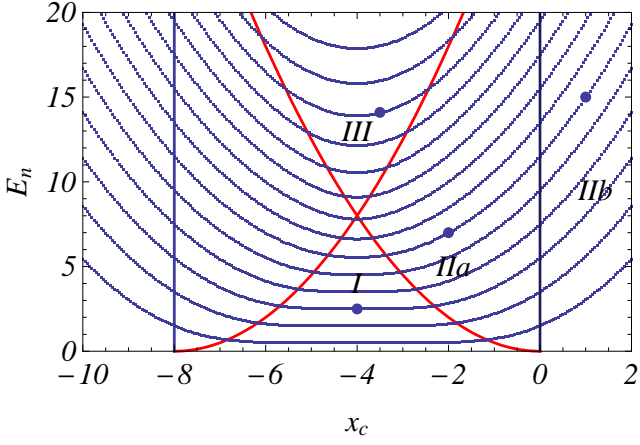} }
\caption{\it Landau levels spectrum calculated numerically for a ribbon of finite width $d=8 \ell_B$. The two vertical lines indicate the position of the edges and the two parabolas indicate the positions $x_c$ for which the classical orbits touch the edges, $x_c=-R$ and $x_c=-d+R$. $x_c$ is written in units of $\ell_B$.}
\label{fig.spectrecomplet_2edges}
\end{figure}

\begin{figure}[!h]
\centerline{ \epsfxsize 6cm \epsffile{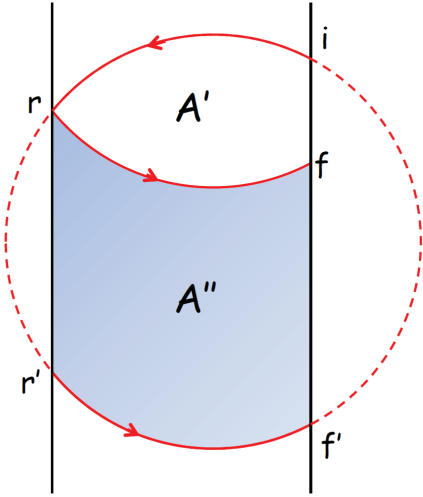} }
\caption{\it In the presence of two walls, the area delimited by a periodic trajectory in the area ${\cal A}'$, but the area  ${\cal A}$ to be quantized is  ${\cal A}={\cal A}'+ {\cal A}''$. }
\label{fig.two-walls}
\end{figure}

\begin{figure}[!h]
\centerline{ \epsfxsize 6cm \epsffile{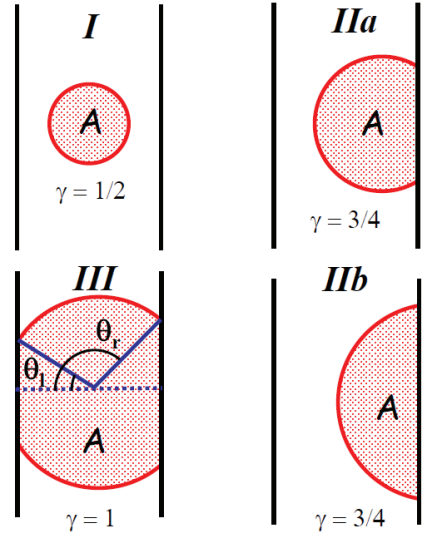} }
\caption{\it Particular trajectories showing the four different possibilities described in Fig. (\ref{fig.spectrecomplet_2edges}). The spectrum is obtained from quantization of the shaded areas. }
\label{fig.trajectoires-particulieres}
\end{figure}

The description of a ribbon with   two edges
is straightforward when the two edges are sufficiently far apart
compared to the cyclotron length. Here we consider the situation where this is not necessarily the case, that is when the distance $d >0$ between the two edges is of the order of a few magnetic lengths $\ell_B$. The spectrum with two edges obtained numerically for $d= 8 \ell_B$ is
shown in Fig. (\ref{fig.spectrecomplet_2edges}), and clearly exhibits three different regions.  We now give a full semiclassical description of this spectrum, considering these three different cases.  Regions I and II have already been discussed and correspond either to a free cyclotron orbit or to an orbit skipping along one boundary. The new interesting case is the region III for which a cyclotron  orbit touches the two boundaries.

We first show that in this case the area $\cal{A}$ to be quantized is the area of a circular orbit cut by the two boundaries (area III in Fig. \ref{fig.trajectoires-particulieres}). This may not seem {\em a priori} obvious since this area is not bounded by a classical trajectory. Actually a periodic trajectory, the arc  $\wideparen{i r f}$  in Fig. \ref{fig.two-walls}, encloses an area $\cal{A}'$ {\em smaller} than ${\cal A}$. Let us return to the argument developed in section (III.A). The Bohr-Sommerfeld quantization rule states that the integral of the velocity along the trajectory  $\wideparen{i r f}$  has to be quantized:

\be m \oint_i^f v_x dx= 2 \pi (n+ \gamma) \hbar  \ . \ee
$i$ and $f$   are respectively the initial and final points of the periodic trajectory and $r$ is  the point where the trajectory bounces on the second boundary (Fig. \ref{fig.two-walls}).
Now the velocity must be calculated with caution. Along the trajectory $\wideparen{i r}$, it is given by $v_x= - \omega_c (y - y_0)$. Then after the bouncing along the second wall, it is now given by   $v_x= - \omega_c (y + y_{rr'} - y_0)$ where $y_{rr'}$ is the distance between the bouncing point $r$ and the point $r'$ which is the next intersection  between the fictitious cyclotron orbit and the second boundary. Therefore we have:

\be {1 \over \omega_c} \oint v_x dx= -\oint_i^f (y-y_0) dx   -\int_r^f y_{rr'} dx \ee

The first integral on the right side is the area delimited by one period of the motion (${\cal A}'$ in Fig. \ref{fig.two-walls}) and the second integral is the shaded area (${\cal A}''$ in Fig. \ref{fig.two-walls}). The sum of these two areas ${\cal A}={\cal A}'+{\cal A}''$ is indeed the total area delimited by the  free cyclotron orbits and the boundaries (Fig. \ref{fig.trajectoires-particulieres}.III).

\begin{figure}[!h]
\centerline{ \epsfxsize 6cm \epsffile{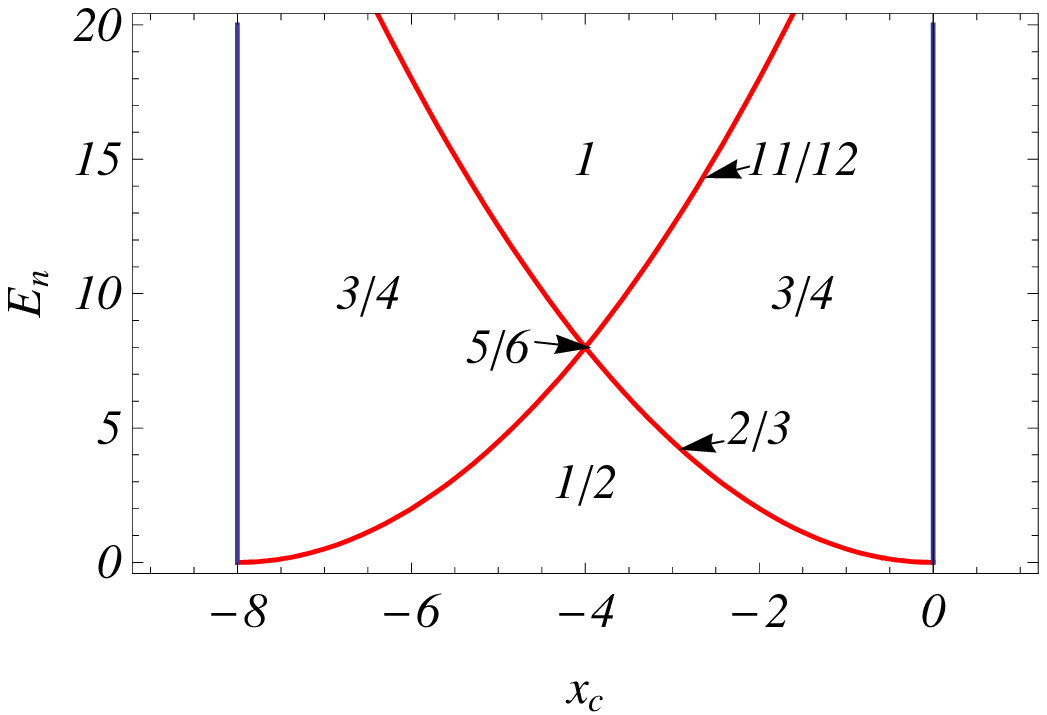} }
\centerline{ \epsfxsize 6cm \epsffile{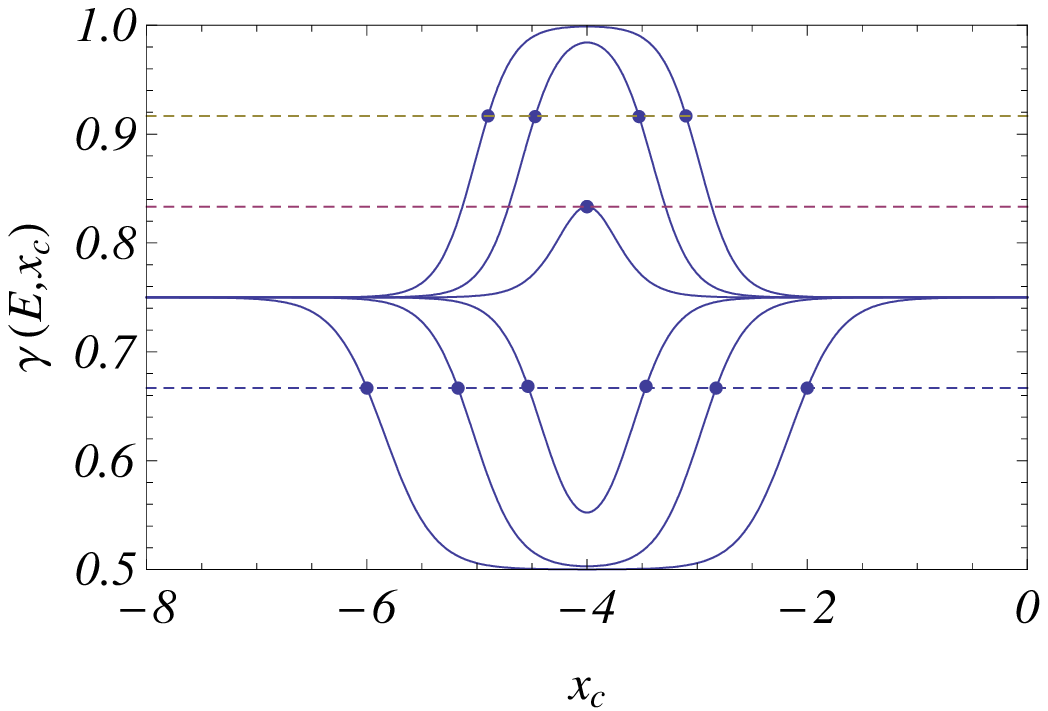} }
\caption{\it Bottom: dependence $\gamma(E,x_c)$ versus $x_c$ for various fixed energies ($E/\hbar \omega_c=2,4,6,8,10,12$, from bottom to top curves). We have indicated the special values   ${1 \over 2}, {2  \over 2},{3 \over 4},{5 \over 6}, {11 \over 12}, 1$  corresponding to the different regions shown on the upper diagram.}
\label{fig.gamma}
\end{figure}

  Defining $d >0 $ as the distance
between the edges, this area  ${\cal A}$ is now given by
\be {\cal A}(E,x_c)= {R^2 \over 2 }[2 \theta_r - \sin 2 \theta_r - 2
\theta_l + \sin 2 \theta_l] \label{Areatheta2edges} \ee
where $\theta_r$ and $\theta_l$ define the position of the cyclotron
orbits with respect to the two edges (see Fig. \ref{fig.trajectoires-particulieres}. We have $\cos \theta_r= x_c/R$
and $\cos \theta_l= (d + x_c)/R$, where $-d<x_c <0$ when the guiding
center is inside the sample.    The energy levels are
semiclassically given by the quantization (\ref{Arean}) of the area ${\cal A}(E,x_c)$ given by (\ref{Areatheta2edges}),
where the  index $\gamma$ depends on the geometry of the orbit. It is the sum of two terms, $\gamma=
\gamma_l+\gamma_r$, where $\gamma_{l,r}=1/4$ in free space,
$\gamma_{l,r}=1/2$ for a skipping orbit, $\gamma_{l,r}=5/12$ when
the cyclotron orbit just touches a wall. Between these different values, $\gamma$ varies continuously. In the general case, we have obtained the value of $\gamma(E,x_c)$ from its decomposition explained above (Eq. \ref{gammadec}). Its value is given  by
\be \gamma= \gamma_l + \gamma_r \qquad \mbox{with} \qquad \gamma_{r,l}={1 \over 4} \ {1 + 4 e^{A X_{r,l}} \over 1 + 2 e^{ A X_{r,l}}} \label{gammarl1} \ee
\begin{eqnarray} X_r&=& \left({2 E \over \hbar \omega_c}\right)^{1/6} (x_c + \sqrt{2 E / \hbar \omega_c}) \nonumber \\
X_l&=& \left({2 E \over \hbar \omega_c}\right)^{1/6} (-x_c-d + \sqrt{{2 E \over \hbar \omega_c}}) \ . \label{gammarl2} \end{eqnarray}
The function $\gamma(E,x_c)$ is shown in Fig. \ref{fig.gamma} as a function of the energy and the position $x_c$ in the ribbon. Note that in the limit where the ribbon is
narrow $d << R$, that is in the high energy regime III, we recover straightforwardly that the area is now
${\cal A}(E,x_c)= 2 d R $, so that the quantization of this area
gives $R= \pi (n+ \gamma) \ell_B^2/d$ and $E_n= { \hbar^2 \over 2 m
} {n'^2 \pi^2 \over d^2}$, with $n'=n+1$, since $\gamma=1$,
corresponding to the two reflections on the edges.

\begin{figure}[!h]
\centerline{ \epsfxsize 8.5cm \epsffile{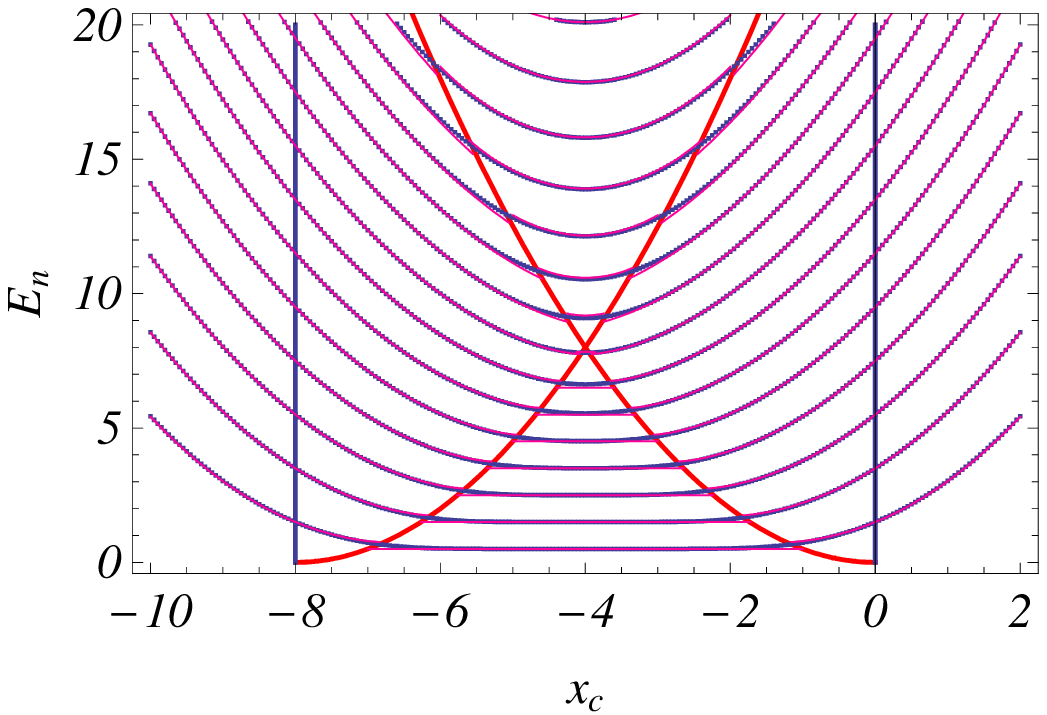} }
\centerline{ \epsfxsize 5cm \epsffile{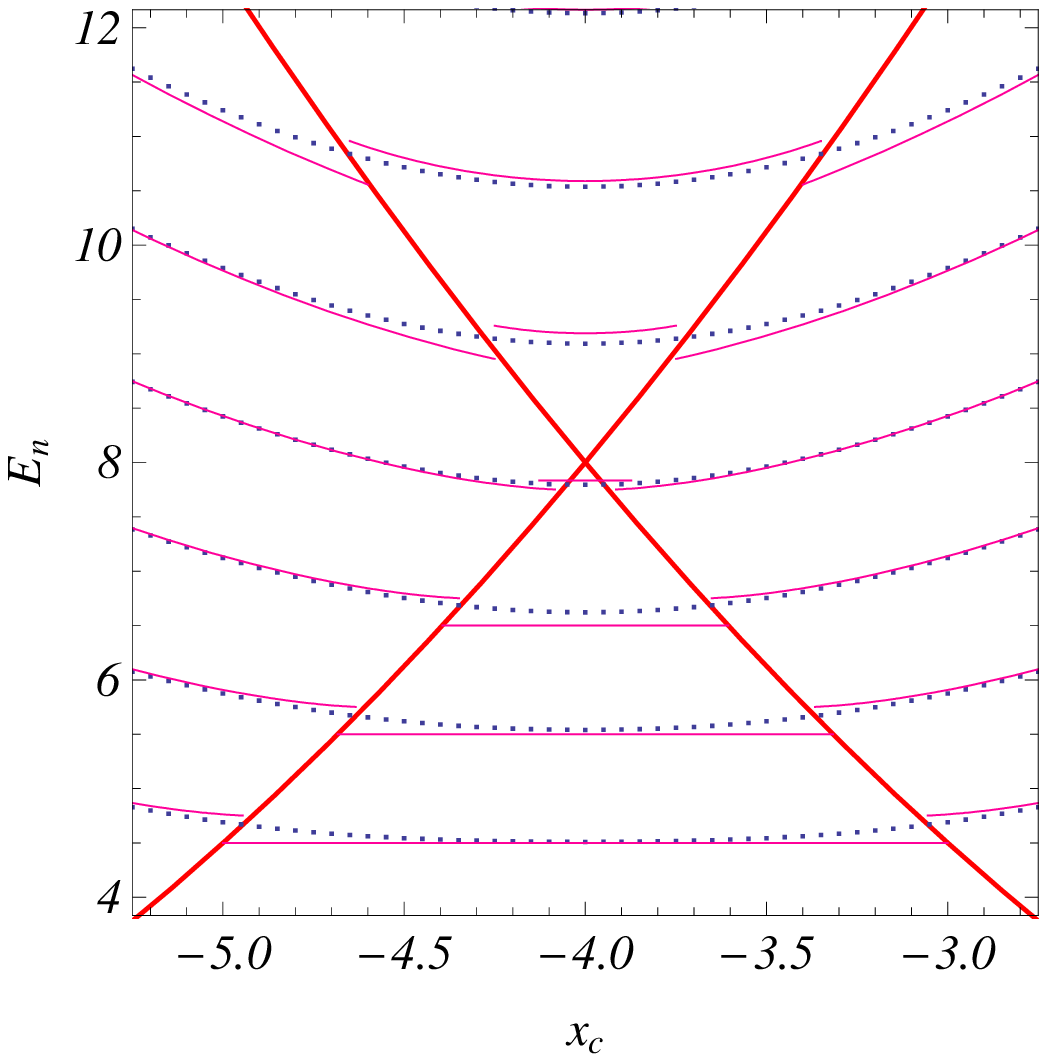} }
\caption{\it Full curves: Spectrum obtained by semiclassical quantization of the area (\ref{Areatheta2edges}), with fixed values of $\gamma=1/2, 3/4, 1$ corresponding respectively to free cyclotron orbits, orbits touching one or two edges. Dotted curves: Exact spectrum obtained by numerical calculation. The semiclassical approximation is quite good except when the classical cyclotron orbit approaches the edges. The two parabolas correspond to $x_c=-R$ and $x_c=-d +R$. }
\label{fig.spectreapproche_2edges}
\end{figure}

\begin{figure}[!h]
\centerline{ \epsfxsize 8.5cm \epsffile{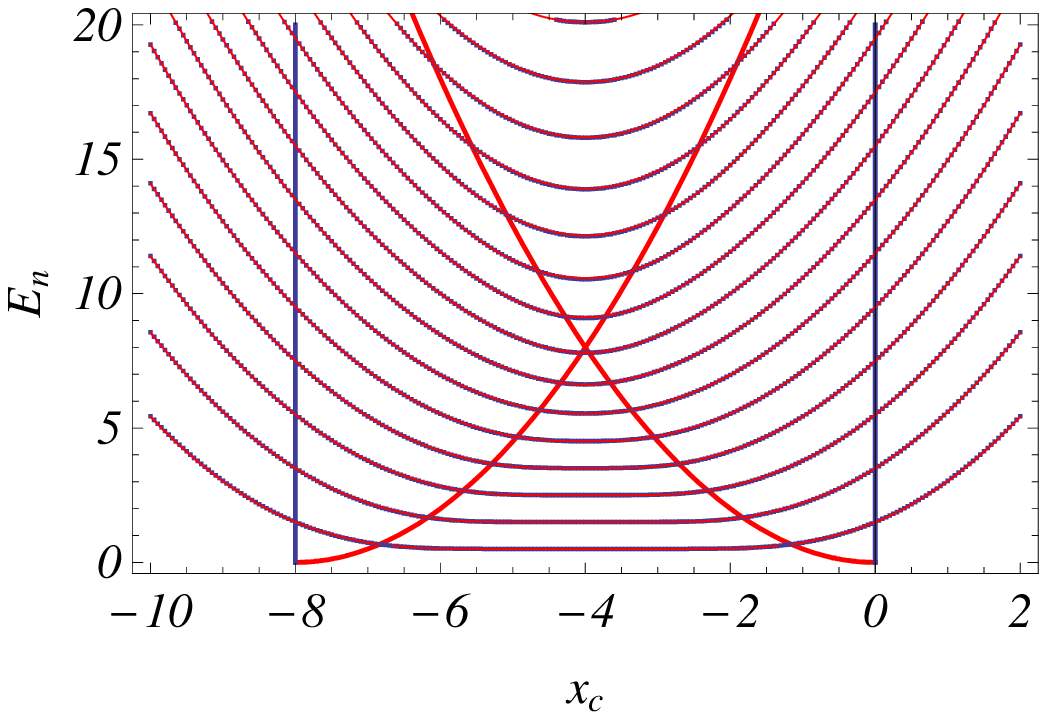} }
\centerline{ \epsfxsize 5cm \epsffile{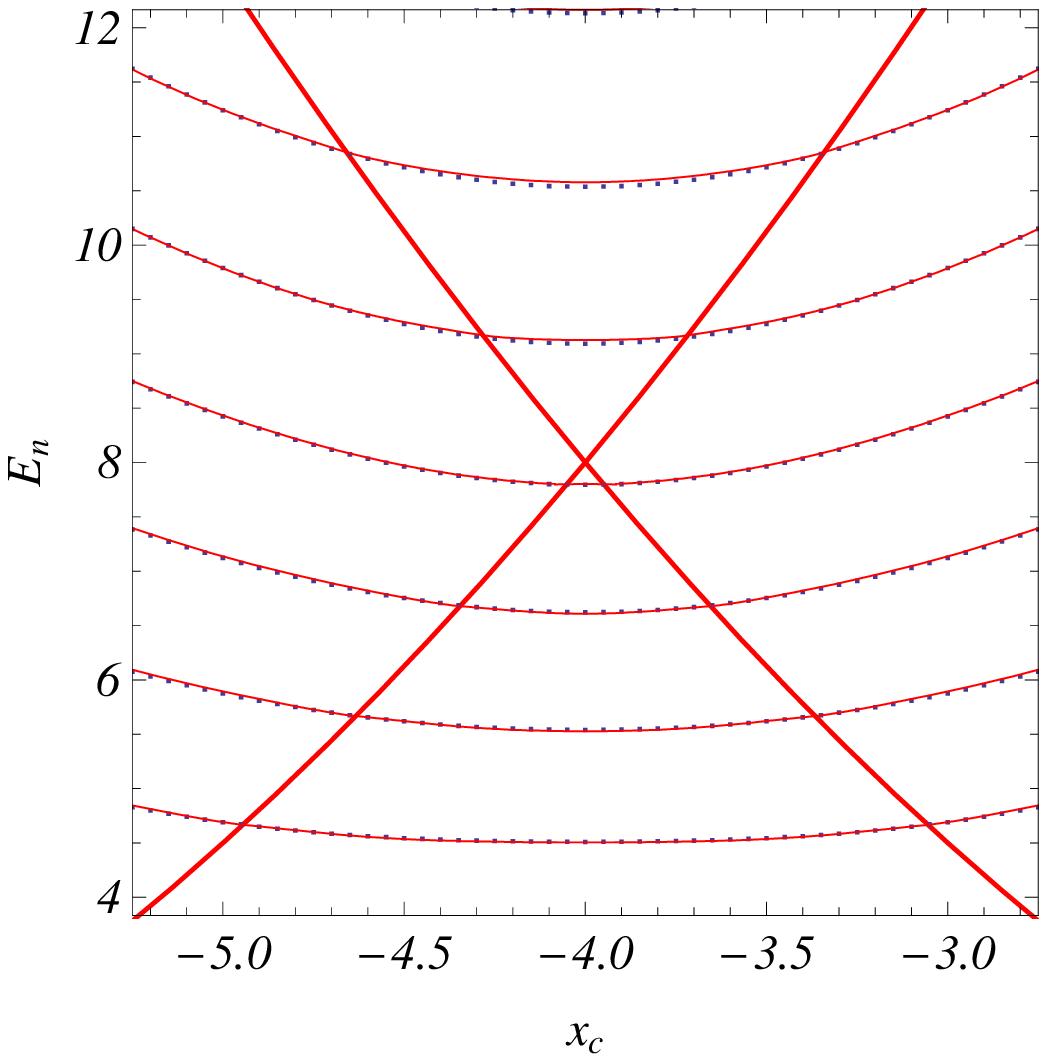} }
\caption{\it Full curves: Spectrum obtained by semiclassical quantization of the area (\ref{Areatheta2edges}), with $\gamma$ given by Eqs. (\ref{gammarl1},\ref{gammarl2}). Dotted curves: Exact spectrum obtained by numerical calculation. This semiclassical WKB approximation is now perfect even  when the classical cyclotron orbit approaches the edges. The two parabolas correspond to $x_c=-R$ and $x_c=-d +R$.}
\label{fig.spectrefitte_2edges}
\end{figure}

The spectrum obtained from semiclassical quantization of the area (\ref{Areatheta2edges}) with fixed values of $\gamma=1/2, 3/4, 1$ corresponding to an free orbit, an orbit touching one or two edges, is displayed in Fig. \ref{fig.spectreapproche_2edges}. The approximation is quite good but there are discontinuities corresponding to $x_c=-R$ and $x_c=-d+R$. In Fig. \ref{fig.spectrefitte_2edges}, the spectrum is obtained from quantization of the area, with the appropriate value of $\gamma$ obtained above (Eqs. \ref{gammarl1},\ref{gammarl2}). We obtain a perfect quantitative description of the full numerical spectrum.

\section{Conclusion}
\label{sect.5}
\begin{figure}[!h]
\centerline{ \epsfxsize 7cm \epsffile{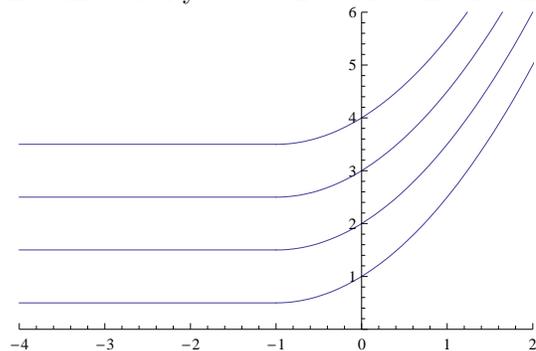} }
\caption{\it Energy levels in the case of a smooth confining
potential $V(x)$. In this case, the energy levels following simply
the potential profile:  $E_n(x_c)= (n+1/2) \hbar \omega_c +
V(x_c)$, compare with Fig. \ref{spectrecomplet}. } \label{fig.smooth-potential}
\end{figure}

We have provided  a semiclassical treatment for the position dependence of the edge states energy levels in the presence of an abrupt infinite potential. This full spectrum may be obtained from the Bohr-Sommerfeld quantization of the area of cyclotron orbits. The orbits do not need to be closed, and the quantization is obtained in all cases, where the orbits hit one or two edges. We provide a simple expression for the mismatch factor $\gamma$ valid for all energies and positions with respect to the boundaries. The situation of an abrupt potential  corresponds to a physical limit where the range of variation of the potential at the edge is much smaller than the magnetic length $\ell_B$. In the case of a smooth potential,  the correct description corresponds to the adiabatic approximation where the energy levels
simply follow the potential $V(x_c)$ at the edge: $E_n(x_c)= (n+1/2)  \hbar \omega_c + V(x_c)$, as shown in Fig. \ref{fig.smooth-potential}. An important difference  with the case of the abrupt potential is that here the energy profile is exactly the same for all levels. In particular, the drift velocity is $independent$ on $n$ and is simply given by
\be v_d= {1 \over e B} {\partial V \over \partial x_c} \ee
and it starts to increase when the potential increases, while for the abrupt potential, the drift velocity depends on $n$, it starts to increase at a distance of order $\sqrt{2 n +1} \ell_B$ from the boundary (Compare Figs. \ref{spectrecomplet} and \ref{fig.smooth-potential}). Another important difference is that, for an abrupt potential, the maximal value reached by the drift velocity is of order of the Fermi velocity $v_E$. If one considers a soft potential of the form $ m \omega^2 (x+x_0)^2/2$, where it is usually assumed  that $\omega < \omega_c$ (this corresponds to the approximation $\ell_B V' < \hbar \omega_c$), the maximal velocity is of order $v_E \omega^2 /\omega_c^2$, much smaller than $v_E$.

In conclusion, we have shown that the semiclassical picture of quantized skipping orbits leads to a
quantitative  description of the edge states energy spectrum. We believe that this quite simple description, not only has a pedagogical interest, but may allow the study of physical quantities not very much discussed in the literature, like the drift velocity. We believe also that it can help for the description of more sophisticated problem like the structure of edge states in graphene.\cite{BreyFertig,Delplace}
 \medskip

\begin{acknowledgements}
 The author thanks J.-N. Fuchs for useful suggestions and comments.
\end{acknowledgements}


\begin{thebibliography}{99}
\bibitem{Halperin82} B. I. Halperin, Phys. Rev. B {\bf 25}, 2185 (1982).
\bibitem{MacDonald84} A. H. Macdonald and P. Streda, Phys. Rev. {\bf B29}, 1616 (1984).
\bibitem{Buttiker88} M. B\"uttiker, Phys. Rev. B {\bf 38}, 9375 (1988).
\bibitem{review} M. B\"uttiker, in {\it Nanostructured systems}, M. Reed ed., Semiconductors and semimetals, {\bf 35}, 191 (Academic Press, Boston, 1991);  C. L. Kane and M. P. A. Fisher, in Perspectives in Quantum Hall Effects, p.109 (Wiley Interscience 2007)
    \bibitem{Beenakker89} H. Van Houten, C.W.J. Beenakker, J.G.
Williamson, M.E. Broekaart, P.H.M. Loosdrecht, B.J. van Wees, J.E.
Mooij, C.T. Foxon and J.J Harris, Phys. Rev. B {\bf 39}, 8556 (1989); H. Van Houten and C.W.J. Beenakker, in {\it Analogies in Optics and Micro Electronics}, W. Van Haeringen and P. Lenstra eds. (Kluwer, Dordrecht, 1990)
\bibitem{BreyFertig}    L. Brey and H. A. Fertig, Phys. Rev. B {\bf 73}, 235411 (2006)
\bibitem{Delplace} P. Delplace and G. Montambaux, in preparation
\bibitem{Maslov} See for example H. Friedrich and J.  Trost, Phys. Rev. A, {\bf 54} (1996). The Maslov indices $\mu_i$ are usually defined such as
$\gamma= (\mu_1 + \mu_2)/4$
\bibitem{Avishai08} Y. Avishai and G. Montambaux, Eur. Phys. J. B, {\bf  66}, 41 (2008)
\bibitem{factor2} In ref. \cite{Avishai08}, the action was defined along half a period.

\end{thebibliography}
\end{document}